\documentclass[a4paper, conference]{IEEEtran}
\IEEEoverridecommandlockouts

\RequirePackage{etex}
\usepackage[pdftex]{graphicx}
\graphicspath{{../pdf/}{../jpeg/}}
\DeclareGraphicsExtensions{.pdf,.jpeg,.png}

\usepackage[cmex10]{amsmath}
\usepackage{algorithmic}
\usepackage{array}
\usepackage{mdwmath}
\usepackage{mdwtab}
\usepackage{url}
\usepackage{tikz}
\usetikzlibrary{matrix,chains,positioning,decorations.pathreplacing,arrows}
\usepackage{mathtools}
\usepackage{booktabs}
\usepackage[hidelinks]{hyperref}
\hypersetup{
    colorlinks=false,
    linkcolor=blue,
    filecolor=magenta,      
    urlcolor=cyan,
}
 
\ifCLASSOPTIONcompsoc
    \usepackage[caption=false,font=footnotesize,labelfont=sf,textfont=sf]{subfig}
\else
    \usepackage[caption=false,font=footnotesize]{subfig}
\fi

\usepackage{cite}
\usepackage{amsmath,amssymb,amsfonts}
\usepackage{algorithmic}
\usepackage{graphicx}
\usepackage{textcomp}
\usepackage{xcolor}
\def\BibTeX{{\rm B\kern-.05em{\sc i\kern-.025em b}\kern-.08em
    T\kern-.1667em\lower.7ex\hbox{E}\kern-.125emX}}

\begin{document}

\onecolumn

\noindent \textcopyright{} 2020 IEEE. Personal use of this material is permitted. Permission from IEEE must be obtained for all
other uses, in any current or future media, including reprinting/republishing this material for advertising or
promotional purposes, creating new collective works, for resale or redistribution to servers or lists, or reuse
of any copyrighted component of this work in other works.

\twocolumn

\title{Composing Graph Theory and Deep Neural Networks to Evaluate SEU Type Soft Error Effects\\
\thanks{This work was supported by the RESCUE ETN project. The RESCUE ETN project has received funding from the European Union's Horizon 2020 Programme under the Marie Skłodowska-Curie actions for research, technological development and demonstration, under grant No. 722325}
}

\author{%
\IEEEauthorblockN{%
  Aneesh Balakrishnan\IEEEauthorrefmark{1}\IEEEauthorrefmark{2},
  Thomas Lange\IEEEauthorrefmark{1}\IEEEauthorrefmark{3},
  Maximilien Glorieux\IEEEauthorrefmark{1},
  Dan Alexandrescu\IEEEauthorrefmark{1},
  Maksim Jenihhin\IEEEauthorrefmark{2}%
}
\IEEEauthorblockA{%
  \IEEEauthorrefmark{1}\textit{iRoC Technologies}, Grenoble, France \\
   \IEEEauthorrefmark{2}\textit{Department of Computer Systems, Tallinn University of Technology}, Tallinn, Estonia \\
  \IEEEauthorrefmark{3}\textit{Dipartimento di Informatica e Automatica, Politecnico di Torino}, Torino, Italy \\
  \{aneesh.balakrishnan, thomas.lange, maximilien.glorieux, dan.alexandrescu\}@iroctech.com \qquad
  maksim.jenihhin@taltech.ee}
}


\maketitle

\begin{abstract}
Rapidly shrinking technology node and voltage scaling increase the susceptibility of Soft Errors in digital circuits. Soft Errors are radiation-induced effects while the radiation particles such as Alpha, Neutrons or Heavy Ions, interact with sensitive regions of microelectronic devices/circuits. The particle hit could be a glancing blow or a penetrating strike. A well apprehended and characterized way of analyzing soft error effects is the fault-injection campaign, but that typically acknowledged as time and resource-consuming simulation strategy. As an alternative to traditional fault injection-based methodologies and to explore the applicability of modern graph based neural network algorithms in the field of reliability modeling, this paper proposes a systematic framework that explores gate-level abstractions to extract and exploit relevant feature representations at low-dimensional vector space. The framework allows the extensive prediction analysis of SEU type soft error effects in a given circuit. A scalable and inductive type representation learning algorithm on graphs called GraphSAGE has been utilized for efficiently extracting structural features of the gate-level netlist, providing a valuable database to exercise a downstream machine learning or deep learning algorithm aiming at predicting fault propagation metrics. Functional Failure Rate (FFR): the predicted fault propagating metric of SEU type fault within the gate-level circuit abstraction of the 10-Gigabit Ethernet MAC (IEEE 802.3) standard circuit.   
\end{abstract}


\section{Introduction}
System engineering focuses on the integration of new small-scale technologies, which constantly advancing the state of the art. Current quality requirements from industrial standards and end-user requirements for high dependability applications expedite reliability modeling and assessment into an increasingly significant endeavor. The aggressive technology node scaling increased the vulnerability of radiation-induced soft errors. The issues due to radiation-based effects, particularly, Single Event Effects (SEEs) seriously impact the circuit's reliability and, the effects of impacts on the functional behavior of the circuit are challenging to evaluate. A valuable approach to tackle the challenge is the fault injection (or) simulation principle that provides precise and accurate information about circuit behaviour under stress and allowing the calculation of actual circuit-level reliability metrics.

\subsection{Motivation}
As mentioned above, the exhaustive fault injection method is the ultimate reliability assessment method in terms of accuracy, but it is very inconvenient in terms of time and EDA licenses; which, makes this approach infeasible on medium and large scale circuits. Therefore, a new test methodology has proposed here. The fundamental idea is to provide an alternate solution to avoid unreasonable test costs by maintaining good statistical significance in results of proposed scope. Research proposals based on Graph Theory and Deep Learning (DL) techniques are more advanced and greatly favoured by researchers to learn statistical dependencies of system-function on related parameters. This motivation develops into a method of applying GraphSAGE algorithm and trying to find the best way to develop relevant feature databases from the gate-level netlist and subsequently applying to a downstream deep neural network for the functional failure reliability metric assessment.

\subsection{Related Works}
 Application of Artificial Intelligence (AI) and Deep Learning approaches to extract feature database of information in the graph network domain, were benefited in different fields. In recent years, different supervised and unsupervised DL approaches have proposed for graphical node embedding. The process of leveraging a node's features into a vector form is called the node embedding. Node2vec \cite{node2vec}, Graph Convolutional Networks (GCN) \cite{kipf2017semi} and GraphSAGE \cite{NIPS2017_6703} have recently gained much attention from researchers for node embedding process. The application of graph-based neural network algorithms (GCN and node2vec) for circuit's reliability modeling, have proposed in papers \cite{8792929} and \cite{8906974} respectively. There is sufficient literature for machine learning (ML) applications in system reliability engineering. But, most of the classical machine learning algorithms rely on black-box modeling (not transparent in modeling the metrics). Here, we aiming a framework which could learn the structural information of circuit's gate-level abstraction in an unsupervised way (without the true target-probability information) based on graphSAGE algorithm and applied these node embedding vectors to a downstream deep learning algorithm. A proper mathematical fault propagating metric given in eq.\ref{SFR} has modeled in this scenario. The analyzed results providing the case of much better numerical superiority in fault propagational metric predictions and interestingly reducing the time complexity. 

\subsection{Organization of the Paper}
The organization of the paper includes a brief introduction followed by sections II, III, IV, and V. Section II covers not only the theoretical background of the physical phenomena and mathematical functions to be modeled but also a brief description of graph theory and graph-based neural networks. The workflow of the framework has provided in section III. Section IV illustrates the results of the fault propagating metric predictions and, finally, a conclusion to the holistic approaches provided in section V. 


\section{Background \& Methodology}
\subsection{Reliability Modeling at Gate-Level}
Single Event Upset (SEU) and Single Event Transient (SET) are the principal consequences of Single Event Effects (SEEs). Single Event Effects are challenging phenomena to analyse or predict when the silicon material of the circuit interacted with the radiation particles. The Single Event Upset widely used here as a prominent SEE representative, and use-case mainly implies an inversion of the stored value in a flip-flop, latch, or memory cell as the result of the radiation-induced charge. Single Event Transient represents a transient pulse of an arbitrary width due to the radioactive event and probably propagate through the combinatorial network and latched to the downstream sequential element. Among SEU and SET events, more probably SEUs will change the state sequences of the circuits and lead to a classified functional failure of the circuits. The functional failure rate due to SEU ($FFR_{i,seu}$) at the given flip-flop ($i$), predicting through this framework. The fault propagational probability metric $FFR_{i,seu}$ described as:     
\begin{align}
\label{SFR}
     FFR_{i,seu} &= FIT_{i,seu} \cdot \prod_{j \in T,L,F}{DR_{i j}} \\
\label{overall_SFR}
     FFR_{seu} &= \sum_{i \in FF} FFR_{i,seu}
\end{align}
where, $DR_{i T}, DR_{i L},\; \text{and} \; DR_{i F}$ represent the fault derating or masking factors such as Temporal Derating (TDR), Logical Derating (LDR) and Functional Derating (FDR) respectively. Similarly, $FIT_{i,seu}$ denotes the rate of soft errors at the flip-flop ($i$) in Failure-In-Time (FIT) unit. Readers could refer the papers \cite{1545891,7086043,6104439,6313869,1175845} for the deep insights about the radiation-induced soft errors and their inevitable intrusive nature in the functioning of microelectronic devices in aggressive radiation environments. 


\subsubsection{\textbf{Temporal Derating}}
    Temporal (or time) derating represents the opportunity window of an event (SET or SEU) and it's probability to be latched to the downstream sequential elements like flip-flop, latch or memory.   

\subsubsection{\textbf{Logical Derating}}
    The porpagational probability of SEU or SET, within the combinational (or) sequential cell networks based on their logical boolean functions is quantified as logical masking probability (or) logical derating factor. 
    
\subsubsection{\textbf{Functional Derating}}
 The probability of the SEU/SET event affects the function of the circuit's actual application. Even though the possibility of changing the circuit's state sequences is significant due to SEU/SET, the effect may be benign or masked because of the application scope.

\subsection{Graph Theory and Deep Learning Algorithms}
\label{Graph Theory and Deep Learning Algorithms}

\subsubsection{\textbf{Graph Theory}}
The graph theory is renowned for a mathematical representation of the data objects and their pairwise relationships in a graph model. In this context, the gate-level abstraction of the circuit has transformed into a graph network where vertices ($\nu$) analogous to the flip-flops and gates, and the directed edges ($\varepsilon$) represent the connection between them from input ports to output ports direction. The mathematical graph-function $G$ of the transformed network given as:
\begin{equation}
\label{Graph:eq}
    G = (\nu,\varepsilon)
\end{equation}

\subsubsection{\textbf{GraphSAGE}}
\label{GraphSAGE}
The GraphSAGE \cite{NIPS2017_6703}, a general inductive framework which leverages node's feature information to efficiently generate node embeddings for previously unseen data. GraphSAGE could be also explained as a graph based neural network with sampler and aggregator functions. Basically the GraphSAGE framework learn a function that generates the node embeddings by sampling and aggregating features from a node's local neighbourhood. Most common approaches like node2vec algorithm \cite{node2vec} require the availability of all the graph-nodes during the training phase of the node embedding process, and those approaches are inherently transductive and generally unable to postulate the learning function to unseen nodes. But an inductive node embedding meant to be an optimized generalization across the graph with same form of features. That is, we can leverage the node features of unseen graph part of a circuit by the embedding generator which trained once with a more generalized graph models of the circuit. This embedding part provides not only the local role of nodes in the graph but also their global positions. A sampler function defines the node's neighborhood definition through a uniform sampling of a fixed number of nodes instead of sampling the entire neighborhood space at each depth-wise iteration. It will result in boosting the optimal usage of memory and reduce run-time complexity. Generally, usage of the word `Depth' means a measure of a fixed distance from the source node for the neighborhood search. At each iteration of depth, an aggregator function has employed. From the state-of-art of the graphSAGE framework, numerous aggregator functions are available like Mean aggregator, Long short-term Memory (LSTM) aggregator, Pooling aggregator and Graph Convoultional Network (GCN) based aggregator. Here we implemented a Pooling aggregator with help of a python neural network libraries. The basic idea of the graphSAGE simplified and explained in the figure \ref{Block_diagram}.      

\subsubsection{\textbf{Deep Neural Network}}
Deep Neural Network (DNN) is an important step in machine learning applications. DNNs are trying to model data of complex distributions by combining different non-linear transformations. The elementary bricks of deep learning approaches are artificial neurons (perceptrons), which are inspired by biological neurons. An artificial neuron combines the input signals with adaptive weights and uses an activation function to deliver the output. In this work, a general fully connected DNN has implemented. The other main categories of deep learning methods are Convolutional Neural Network (CNN) and Recurrent Neural Networks (RNN).

\subsection{Fault Injection Simulation Paradigm}
\begin{figure}[ht]
    \centering
    \includegraphics[width=\linewidth]{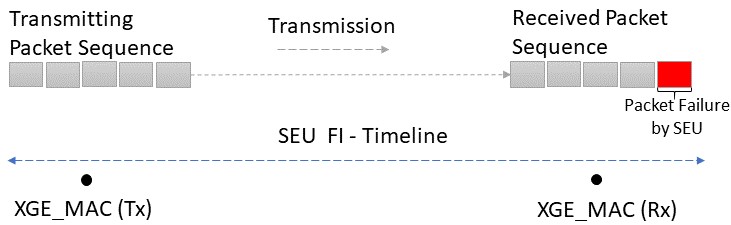}
    \caption{Transmission between XGE\_MAC Transmitter (TX) and it's Receiver (RX) }
    \label{Test_paradigm}
\end{figure}

As above mentioned, the true database required to train and predict the fault propagating metric ($FFR_{i,seu}$), obtained through an exhaustive Fault Injection (FI) campaign. In an exhaustive FI campaign, an SEU type fault injected independently at each flip-flop in each clock cycle of the time duration between transmission and reception of the input packets, as given in figure \ref{Test_paradigm}. If the injected fault (SEU) in a single clock cycle propagates through the circuit and subsequently causes the circuit's function to fail, it will account for the functional failure. Finally, $FFR_{i,seu}$ was obtained by summing the functional failures per flip-flop over the total number of clock cycles required for the operation. In total, 1100 flip-flops from different blocks of the circuit (such as TX, RX, Wishbone Interface, Fault State-machine, and Sync\_clk), are tested and recorded functional failure rates ($FFR_{i,seu}$) as true database. 

\section{Methodology Illustration}
\begin{figure*}[ht]
    \centering
    \includegraphics[width=0.95\linewidth]{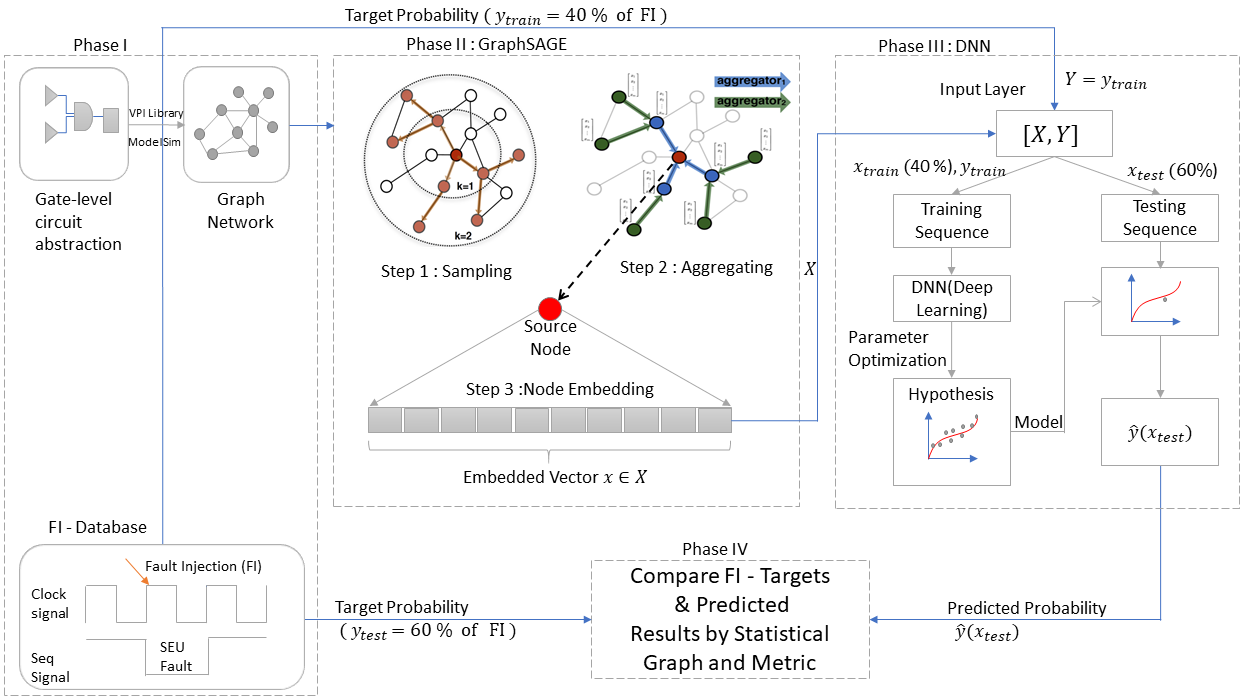}
    \caption{A Systematic workflow diagram of the implemented scientific work}
    \label{Block_diagram}
\end{figure*}

The whole approach has explained through successive work-phases as follows: 
\subsection{Phase I}
The proposed work implements a method to map the gate-level netlist as a Probabilistic Bayesian Graph (PGB) in the Graph Modeling Language (GML) format. To accomplish this goal, a Verilog Procedural Interface (VPI) based library function (a user-defined library) linked to a standard simulation tool (ModelSim/open-source tool). A gate-level netlist mapped to the graph model has represented in figure \ref{Block_diagram}. Parallelly, the FI database simulated in this phase of the workflow. 
\subsection{Phase II}
In the second phase of the approach, a feature matrix ($X$) corresponding to graph nodes extracted using the GrpahSAGE algorithm. As mentioned in section \ref{GraphSAGE}, GraphSAGE includes two principal steps. The premier step was the sampler algorithm. The sampler algorithm defines the neighbourhood space of a source node. In this scenario, we defined the parameter $K = 2$, which means that the sampler will sample up to the depth of 2 neighbourhood space. In the second step of the GraphSAGE algorithm, an aggregator has implemented at each depth ($1 \leq k \leq K$). This could be seen in Phase II of figure \ref{Block_diagram}, where blue and green line indicates the aggregators at depth $k=1$ and $k=2$ respectively. Here, a max-pooling aggregator was implemented. The mathematical abstraction of the pooling aggregator \cite{NIPS2017_6703} formulated as:

\begin{equation}
\label{Aggregator}
AGGRE_{k}^{pool} = max( \{ \sigma(W_{pool}h_{u_i}^{k} + b), \forall{u_i}\in N_k(v)\}),
\end{equation}

where equation \ref{Aggregator} represents the aggregator function at depth $k$ and it basically a neural network with parameters $W_{pool}$ and $b$. Parameters optimized through unsupervised learning. $N_k(v)$ represents k-neighbourhood of vertex $v$ and $h_{u_i}^{k}$ indicates the aggregated neighborhood vector and, $\sigma$ is the activation function of the neural network. In this way, we could represent the whole GraphSAGE algorithm as a graph-based neural network. At the end of this phase, each node reformed into a corresponding vector and alternatively form a matrix representation ($X$) of the circuit as given in figure \ref{Block_diagram}. 

\subsection{Phase III}
Phase III of figure \ref{Block_diagram} elucidates the DNN algorithm that exercised for prediction purposes. There are two parts included in phase III. The first part is the training part of DNN, and the second one is the testing part of DNN. In the training part, 40 \% of the feature matrix and corresponding target probability metric from FI - database, are taken to postulate a hypothesis that best describes the target probability distribution ($FFR_{i,seu}$) by supervised learning method. The optimized parameters (Weights and Bias) of the best fit of the target distribution should provide as model parameters. In the testing part, the proposed model applies to an unknown input vector and predicts the target probability metric. The DNN architecture consists of 5 dense layers, including the input and the output layers. 

\subsection{Phase IV}
The final phase includes a comparison between the predicted and target probability metrics. The compared results plotted in figure \ref{Result:1}, as well as the impacts of results provided in table \ref{Impact Table}.


\section{Experimental Results}
\begin{figure}[ht]
    \centering
    \includegraphics[width=1\linewidth]{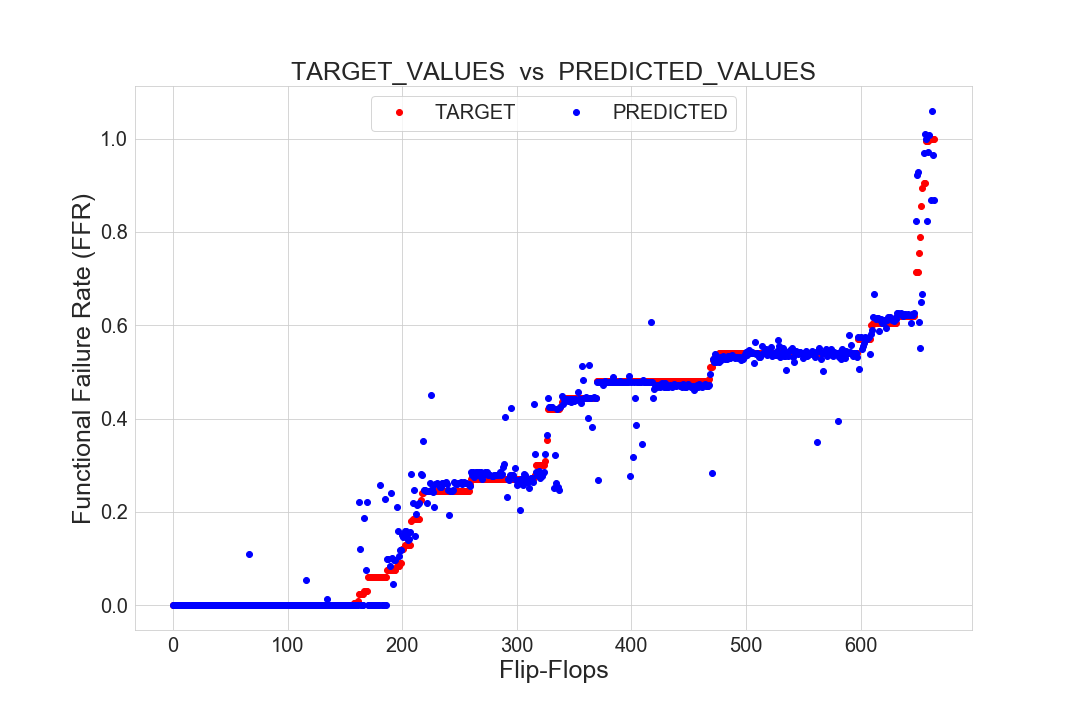}
    \caption{Graph Comparison : Mean Absolute Error (MAE) = 0.0186 and coefficient of determination ($R^2$ Metric) = 0.96 with a test size of 60\%}
    \label{Result:1}
\end{figure}

In figure \ref{Result:1}, the functional failure rates of flip-flops to be predicted were shown in red color, and the corresponding predicted probabilities have shown in blue color. The case study has conducted with the gate-level circuit of the 10-Gigabit Ethernet MAC. The graphical comparison gives a visualization of how well the prediction replicates the observed database. In this case, the DNN prediction achieves the coefficient of determination ($R^2$) value of approximately 0.96, where the best model fit value of $R^2$ metric is 1, and the worst value is 0. In statistics, the R-squared ($R^2$) value is the measure of goodness-of-fit of the proposed regression model and, the projected R-squared value (0.96) able to explain most of the variation in the response data. The entire work repeated and achieves a good prediction accuracy with other standard circuits (e.g., The USB 1.1 Function IP Core).

\begin{table}[ht]
    \centering
    \caption{Impact of Prediction in Time and Tool Requirements}
    \label{Impact Table}
    \begin{tabular}{l c c c}
    \toprule
        Model & Time & Tool & Model Fit ($R^2$) \\
    \midrule
        Fault Injection & 17 hours & 7 Modelsim & Target Model \\
        Fault Injection & $\approx$ 5 days & 1 Modelsim & Target Model \\
        \hline
        \\[\dimexpr-\normalbaselineskip+2pt]
        \begin{tabular}{@{}l@{}}
                   GraphSAGE + DNN  \\
                   (Test + Training) \\
        \end{tabular} & $<$ 10 minutes & 1 Modelsim & 0.96 \\
        
    \bottomrule
    \end{tabular}
\end{table}

Table \ref{Impact Table} outlines the impacts of accelerated predictions in terms of time and simulation tool requirements. Even-though GraphSAGE and DNN based ensemble algorithm provide a significant reduction in the required test resources without compromising the quality of modeling, the implemented algorithm depends on 40\% of FI-database for training the downstream DNN as picturized in Phase III of fig.\ref{Block_diagram}. But, it is quite impressive to note that the test and training phase of the whole algorithm takes only less than 10 minutes.

\section{Conclusion}
An accelerated testing methodology; that is scalable and very cost-effective in resource handling, has developed for medium and largescale circuits to predict Functional Failure Rate due to SEU type fault without dropping the significance of the statistical modeling. 

\bibliographystyle{IEEEtran}
\bibliography{BIBLIOGRAPHY/IEEE_MECO_2020}

\begin{thebibliography}{10}
\providecommand{\url}[1]{#1}
\csname url@samestyle\endcsname
\providecommand{\newblock}{\relax}
\providecommand{\bibinfo}[2]{#2}
\providecommand{\BIBentrySTDinterwordspacing}{\spaceskip=0pt\relax}
\providecommand{\BIBentryALTinterwordstretchfactor}{4}
\providecommand{\BIBentryALTinterwordspacing}{\spaceskip=\fontdimen2\font plus
\BIBentryALTinterwordstretchfactor\fontdimen3\font minus
  \fontdimen4\font\relax}
\providecommand{\BIBforeignlanguage}[2]{{%
\expandafter\ifx\csname l@#1\endcsname\relax
\typeout{** WARNING: IEEEtran.bst: No hyphenation pattern has been}%
\typeout{** loaded for the language `#1'. Using the pattern for}%
\typeout{** the default language instead.}%
\else
\language=\csname l@#1\endcsname
\fi
#2}}
\providecommand{\BIBdecl}{\relax}
\BIBdecl

\bibitem{node2vec}
A.~Grover and J.~Leskovec, ``node2vec: Scalable feature learning for
  networks,'' in \emph{ACM SIGKDD International Conference on Knowledge
  Discovery and Data Mining (KDD)}, 07 2016, pp. 855--864.

\bibitem{kipf2017semi}
T.~N. Kipf and M.~Welling, ``Semi-supervised classification with graph
  convolutional networks,'' in \emph{International Conference on Learning
  Representations (ICLR)}, 2017.

\bibitem{NIPS2017_6703}
\BIBentryALTinterwordspacing
W.~Hamilton, R.~Ying, and J.~Leskovec, ``Inductive representation learning on
  large graphs,'' in \emph{Advances in Neural Information Processing Systems
  30}, 2017, pp. 1024--1034. [Online]. Available:
  \url{http://papers.nips.cc/paper/6703-inductive-representation-learning-on-large-graphs.pdf}
\BIBentrySTDinterwordspacing

\bibitem{8792929}
A.~{Balakrishnan}, T.~{Lange}, M.~{Glorieux}, D.~{Alexandrescu}, and
  M.~{Jenihhin}, ``Modeling gate-level abstraction hierarchy using graph
  convolutional neural networks to predict functional de-rating factors,'' in
  \emph{2019 NASA/ESA Conference on Adaptive Hardware and Systems (AHS)}, 2019,
  pp. 72--78.

\bibitem{8906974}
A.~{Balakrishnan}, T.~{Lange}, M.~{Glorieux}, D.~{Alexandrescu}, and
  M.~{Jenihhin}, ``The validation of graph model-based, gate level
  low-dimensional feature data for machine learning applications,'' in
  \emph{2019 IEEE Nordic Circuits and Systems Conference (NORCAS): NORCHIP and
  International Symposium of System-on-Chip (SoC)}, 2019, pp. 1--7.

\bibitem{1545891}
R.~C. {Baumann}, ``Radiation-induced soft errors in advanced semiconductor
  technologies,'' \emph{IEEE Transactions on Device and Materials Reliability},
  vol.~5, no.~3, pp. 305--316, 2005.

\bibitem{7086043}
M.~{Ebrahimi}, A.~{Evans}, M.~B. {Tahoori}, E.~{Costenaro}, D.~{Alexandrescu},
  V.~{Chandra}, and R.~{Seyyedi}, ``Comprehensive analysis of sequential and
  combinational soft errors in an embedded processor,'' \emph{IEEE Transactions
  on Computer-Aided Design of Integrated Circuits and Systems}, vol.~34,
  no.~10, pp. 1586--1599, 2015.

\bibitem{6104439}
D.~{Alexandrescu}, E.~{Costenaro}, and M.~{Nicolaidis}, ``A practical approach
  to single event transients analysis for highly complex designs,'' in
  \emph{2011 IEEE International Symposium on Defect and Fault Tolerance in VLSI
  and Nanotechnology Systems}, 2011, pp. 155--163.

\bibitem{6313869}
D.~{Alexandrescu} and E.~{Costenaro}, ``Towards optimized functional evaluation
  of see-induced failures in complex designs,'' in \emph{2012 IEEE 18th
  International On-Line Testing Symposium (IOLTS)}, 2012, pp. 182--187.

\bibitem{1175845}
R.~{Baumann}, ``The impact of technology scaling on soft error rate performance
  and limits to the efficacy of error correction,'' in \emph{Digest.
  International Electron Devices Meeting,}, 2002, pp. 329--332.

\end{thebibliography}

\end{document}